\documentclass[aps,reprint,prl,superscriptaddress,showpacs,floats,floatfix]{revtex4-1}
\usepackage{graphicx}
\usepackage{amsmath,graphicx,dcolumn}
\usepackage{hyperref}
\usepackage[usenames]{color}
\usepackage{datetime}
\usepackage{color}
\usepackage{textcomp}
\usepackage{ulem}

\newcommand*{\CPI}{CePt$_2$In$_7$}

\usepackage{graphicx}

\begin{document}

\title{Angle-resolved photoemission spectroscopy study of crystal electric field in heavy fermion compound {\CPI}}

\author{Yang Luo}
\affiliation{School of Physics and Electronics, Central South University, Changsha 410083, Hunan, Peoples Republic of China}

\author{Chen Zhang}
\affiliation{School of Physics and Electronics, Central South University, Changsha 410083, Hunan, Peoples Republic of China}

\author{Qi-Yi Wu}
\affiliation{School of Physics and Electronics, Central South University, Changsha 410083, Hunan, Peoples Republic of China}

\author{Fan-Ying Wu}
\affiliation{School of Physics and Electronics, Central South University, Changsha 410083, Hunan, Peoples Republic of China}

\author{Jiao-Jiao Song}
\affiliation{School of Physics and Electronics, Central South University, Changsha 410083, Hunan, Peoples Republic of China}

\author{W. Xia}
\affiliation{School of Physical Science and Technology, ShanghaiTech University, Shanghai 201210, Peoples Republic of China}
\affiliation{School of Physical Sciences, University of Chinese Academy of Sciences, Beijing 100049, China}

\author{Yanfeng Guo}
\affiliation{School of Physical Science and Technology, ShanghaiTech University, Shanghai 201210, Peoples Republic of China}
\affiliation{School of Physical Sciences, University of Chinese Academy of Sciences, Beijing 100049, China}

\author{J\'{a}n Rusz}
\affiliation{Department of Physics and Astronomy, Uppsala University, Box 516, S-75120 Uppsala, Sweden}

\author{Peter M. Oppeneer}
\affiliation{Department of Physics and Astronomy, Uppsala University, Box 516, S-75120 Uppsala, Sweden}

\author{Tomasz Durakiewicz}
\affiliation{Institute of Physics, Maria Curie Sklodowska University, 20-031 Lublin, Poland}

\author{Yin-Zou Zhao}
\affiliation{School of Physics and Electronics, Central South University, Changsha 410083, Hunan, Peoples Republic of China}

\author{Hao Liu}
\affiliation{School of Physics and Electronics, Central South University, Changsha 410083, Hunan, Peoples Republic of China}

\author{Shuang-Xing Zhu}
\affiliation{School of Physics and Electronics, Central South University, Changsha 410083, Hunan, Peoples Republic of China}

\author{Ya-Hua Yuan}
\affiliation{School of Physics and Electronics, Central South University, Changsha 410083, Hunan, Peoples Republic of China}

\author{Jun He}
\affiliation{School of Physics and Electronics, Central South University, Changsha 410083, Hunan, Peoples Republic of China}

\author{Shi-Yong Tan}
\affiliation{Science and Technology on Surface Physics and Chemistry Laboratory, Mianyang 621908, Peoples Republic of China}

\author{Y. B. Huang}
\affiliation{Shanghai Institute of Applied Physics, CAS, Shanghai, 201204, China, Peoples Republic of China}

\author{Zhe Sun}
\affiliation{National Synchrotron Radiation Laboratory, University of Science and Technology of China, Hefei, Peoples Republic of China}

\author{Yi Liu}
\affiliation{National Synchrotron Radiation Laboratory, University of Science and Technology of China, Hefei, Peoples Republic of China}

\author{H. Y. Liu}
\affiliation{Beijing Academy of Quantum Information Sciences, Beijing 100085, China}

\author{Yu-Xia Duan}
\affiliation{School of Physics and Electronics, Central South University, Changsha 410083, Hunan, Peoples Republic of China}

\author{Jian-Qiao Meng}
\email{Corresponding author: jqmeng@csu.edu.cn}\affiliation{School of Physics and Electronics, Central South University, Changsha 410083, Hunan, Peoples Republic of China}
\affiliation{Synergetic Innovation Center for Quantum Effects and Applications (SICQEA), Hunan Normal University, Changsha 410081, Peoples Republic of China}

\date{\today}

\begin{abstract}
The three-dimensional electronic structure and Ce 4$f$ electrons of the heavy fermion superconductor {\CPI} is investigated. Angle-resolved photoemission spectroscopy using variable photon energy establishes the existence of quasi-two and three dimensional Fermi surface topologies. Temperature-dependent 4$d$-4$f$ on-resonance photoemission spectroscopies reveal that heavy quasiparticle bands begin to form at a temperature well above the characteristic (coherence) temperature $T$*. $T$* emergence may be closely related to crystal electric field splitting, particularly the low-lying heavy band formed by crystal electric field splitting.
\end{abstract}

\pacs{74.25.Jb,71.18.+y,74.70.Tx,79.60.-i}
\maketitle

Heavy fermions (HF) have long been the focus of condensed matter physics due to their rich and interesting physical phenomena \cite{Gegenwart2008, Pfleiderer2009} which includes quantum criticality, magnetic order and unconventional superconductivity coexistence, and non-Fermi liquid behaviors. This system's diverse and tunable ground states result in various materials types, such as HF superconductors, Kondo insulators, and HF antiferromagnetism (AFM) or ferromagnetism (FM). HFs ground state can be easily tuned using pressure, magnetic field, and doping \cite{Thompson2012}.

A characteristic temperature, $T$*, of HF materials, called the ``hybridization temperature" or the ``coherent temperature" is commonly defined. There are different views on the physics which underlies $T$* and what determines it. $T$* is currently believed to be collective hybridization onset between incoherent, localized $f$ electrons and conduction band electrons \cite{YPLiu2019}. This leads to the emergence of heavy electrons at lower temperatures \cite{YFYang2008, YFYang2016}. Several techniques are available to determine $T$*, including such as thermodynamics, transport, Knight shift measurements \cite{Warren2011}, nuclear magnetic resonance (NMR) \cite{HSakai2012}, and inelastic neutron scattering \cite{TWillers2010}. A question that remains is, whether any $T$* determined by various techniques, or researchers using the same techniques, has the same underlying physics. For example, inelastic neutron scattering studies have revealed that the characteristic temperature $T$* of Ce$M$In$_5$ is tens of K \cite{TWillers2010}, angle-resolved photoemission spectroscopy (ARPES) studies have found that heavy quasiparticle bands begin to form well above $T$*, and show a crossover behavior across $T$* \cite{QYChen2017,QYChen2018A, QYChen2018B, QYao2019}. A laser-based ARPES study of YbRh$_2$Si$_2$ shows that coherent states developed just below $T$* \cite{SKMo2012}. Knowing what determines $T$* is of fundamental and practical importance and is a prerequisite for understanding HF physics.

We chose the AFM HF superconductor {\CPI} to study this concern. {\CPI}, discovered in 2008 \cite{Kurenbaeva2008}, is a member of a widely-studied HF compound family Ce$_m$$M$$_n$In$_{3m+2n}$ ($M$ = Co, Rh, Ir, Pt)\cite{QYChen2017, QYChen2018A, QYChen2018B, QYao2019, Fujimori2003, Fujimori2006, HJLiu2019, FSteglich1991, RSettai2007, Fujimori2016, RJiang2015, YXDuan2019}. At ambient pressure {\CPI} undergoes an AFM phase transition at the N\'{e}el temperature $T$$_N$ = 5.2 K \cite{Tobash2012}. Under pressure, {\CPI} has a bulk superconducting transition temperature of up to $T$$_c$ =2.1 K (near 3.5 GPa \cite{EDBauer2010A}), AFM and superconductivity exist simultaneously within a certain pressure range \cite{EDBauer2010B, EDBauer2010A, HSakai2014}. Nuclear quadrupolar resonance (NQR) and muon spin rotation/relaxation data reveal commensurate \cite{Warren2010, MMansson2014}, or coexisting commensurate and incommensurate \cite{HSakai2011, HSakai2012}, AFM orders. A complex Fermi-surface (FS) topology composed of quasi-two-dimensional (quasi-2D) sheets or coexisting quasi-2D and three-dimensional (3D) character were revealed by quantum oscillation \cite{MMAltarawneh2011, KGotze2017} and ARPES measurements \cite{BShen2017, YXDuan2019}. Recently, weaker hybridization was revealed by optical spectroscopy \cite{RYChen2016} and ARPES \cite{YXDuan2019}. Knight shift measurements return a $T$* $\sim$ 40 K \cite{Warren2011}, and NMR determines a $T$* $\sim$ 20 K \cite{HSakai2012}.

\begin{figure}[t!]
\vspace*{-0.2cm}
\begin{center}
\includegraphics[width=0.95\columnwidth,angle=0]{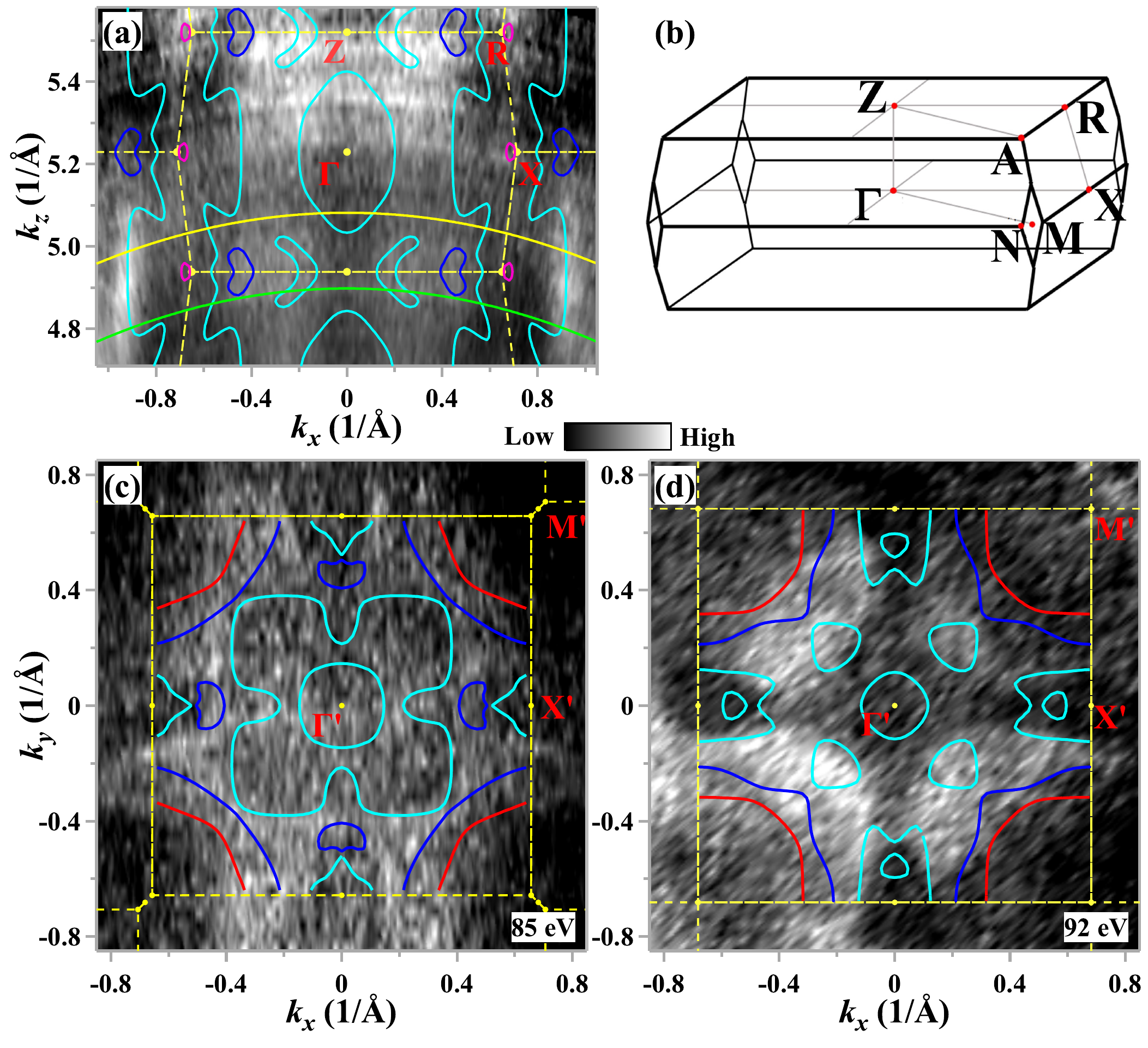}
\end{center}
\vspace*{-0.7cm}
\caption{(color online) {\CPI} FS at 20 K. ({\bf {a}}) Experimental 3D FS maps measured using $h$$\nu$ = 76-112 eV photons in 1 eV steps, in the $\Gamma$$Z$$R$$X$ plane. ({\bf {b}}) A 3D Brillouin zone (BZ) of {\CPI} with high-symmetry momentum points (red dots) marked. ({\bf {c}}) and ({\bf {d}}) Spectral weight as a function of 2D momentum ($k$$_x$, $k$$_y$) taken with 85 and 92 eV photons, respectively. Momentum cuts with 85 and 92 eV photons are marked with green and yellow lines, respectively, in (a). All photoemission intensity data were integrated over an [-20 meV, 20 meV] energy window with respect to the Fermi energy $E$$_F$. BZ and calculated results are adopted from reference \cite{YXDuan2019}}
\end{figure}

In this study we varied the temperature  between 10 and 136 K above the N\'{e}el temperature. {\CPI} FS topology along the $k$$_z$ (perpendicular) direction was mapped out using systematic photon energy dependence ($h$$\nu$ = 76 - 112 eV) and constant photon energy ($h$$\nu$ = 85 and 92 eV) ARPES measurement. The FS measured topology was compared to density-functional theory (DFT) calculations \cite{YXDuan2019}. The 4$f$ electrons properties were investigated by comparing off- and on-resonance spectra. Temperature-dependent studies showed that below 60 K, crystal electric field (CEF) splitting resulted in ``relocalization" of the itinerant $f$ electrons.

High-quality {\CPI} single crystals were grown using an In self-flux method. Data presented in Figs.\ 1(a) and 2 were obtained at the ``Dreamline" beamline of the Shanghai Synchrotron Radiation Facility (SSRF) using a Scienta DA30 analyzer, and the vacuum was kept below 1$\times$10$^{-10}$ mbar. Data shown in Figs.\ 1(c, d), 3, and 4 were obtained at beamline 5-2 of the Stanford Synchrotron Radiation Lightsource (SSRL) using a Scienta D80 analyzer, with a base pressure of better than 6 $\times$ 10$^{-11}$ mbar. Typical angular resolution was $\sim$0.2$^{\circ}$. The overall energy resolution is better than 20 meV. All samples were cleaved $in$ $situ$ at $\sim$ 15 K. On-resonance 121/123 eV photons and off-resonance 114 eV photons were used to investigate the nature of Ce 4$f$ electrons. Both $s-$ and $p-$ polarized photons were used. The $s-$polarized light is perpendicular to the plane defined by the incident light and emitted electrons. The $p-$polarized light is parallel to this plane.

\begin{figure}[tb!]
\vspace*{-0.2cm}
\begin{center}
\includegraphics[width=0.95\columnwidth,angle=0]{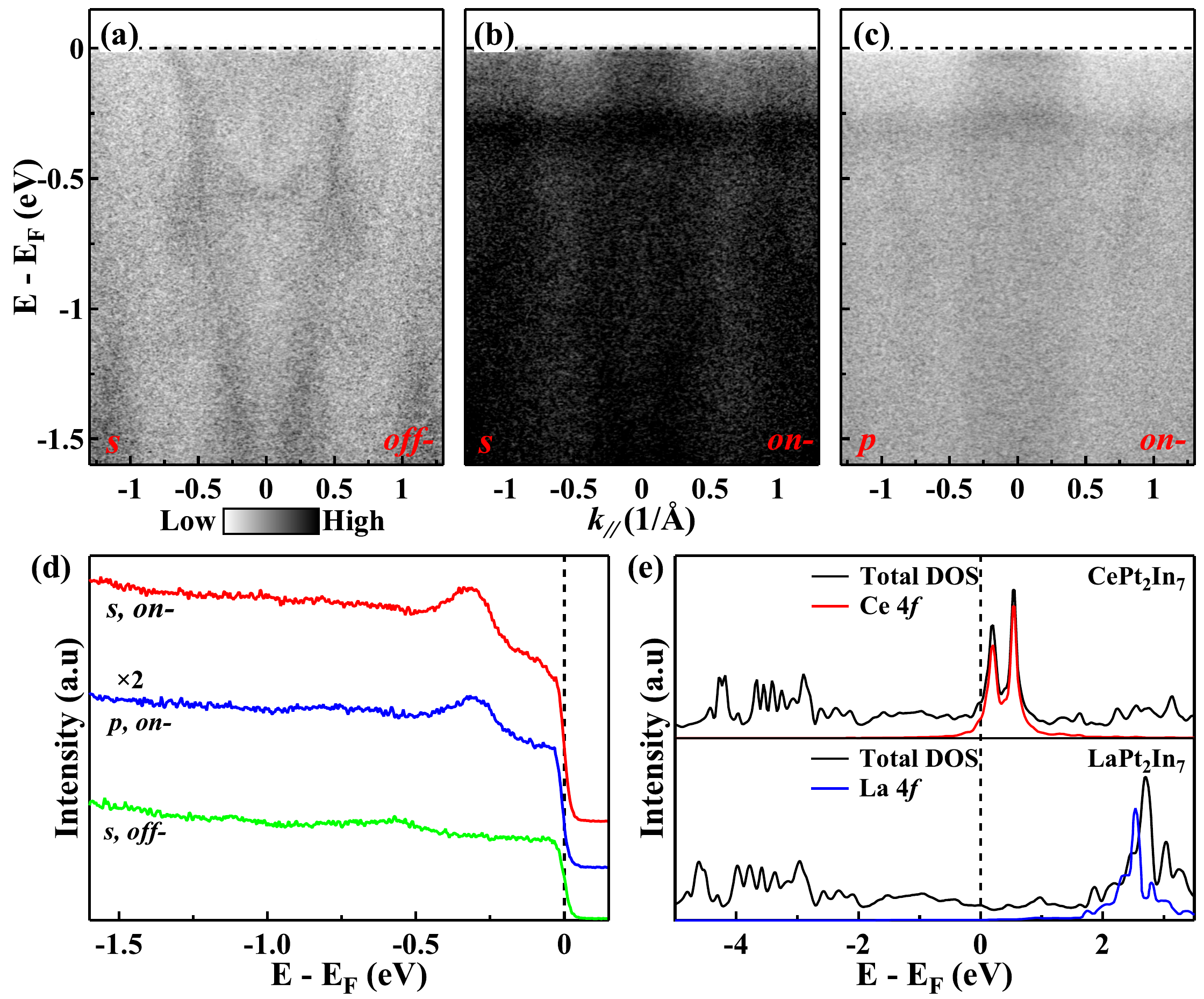}
\end{center}
\vspace*{-0.7cm}
\caption{(color online) On- and off-resonance ARPES data for {\CPI} along X-$\Gamma$-X taken at 20 K with ({\bf {a}}) off-resonance (114 eV) $s$-polarized, ({\bf {b}}) on-resonance (123 eV) $s$-polarized, and ({\bf {c}}) on-resonance $p$-polarized light. ({\bf {d}}) Angle-integrated photoemission spectroscopy of the intensity plot in (a)-(c). The integrated momentum range is [-0.15 {\AA}$^{-1}$, 0.15 {\AA}$^{-1}$]. ({\bf {e}}) Comparison of density of state (DOS) vs energy $E$ for the isostructural compounds {\CPI} (upper panel, adopted from reference \cite{YXDuan2019}) and LaPt$_2$In$_7$ (lower panel).}
\end{figure}

 {\CPI} FS topologies were obtained from photon-energy-dependent normal emission at a temperature of 15 K (Fig.\ 1(a)). The measurement was performed in a section of the high-symmetry $\Gamma$$Z$$R$$X$ plane of the {\CPI} BZ. Different $k$$_z$ values were accessed by varying photon energies between 76 and 112 eV. The corresponding $k$$_z$ range covers more than a BZ and included both $\Gamma$ and $Z$ points. DFT calculated Fermi contours \cite{YXDuan2019} were overlaid on measured intensities. Fermi sheets intensity and shapes vary with photon energy. This demonstrates the 3D character of the electronic structure of {\CPI} along the $\Gamma$-$X$ direction. This is consistent with theoretical calculations \cite{YXDuan2019, BShen2017, Klimczuk2014}. A strong $k$$_z$ dispersion in the data allows us to determine an approximate inner potential $V$$_0$ $\sim$ 11 eV, consistent with prior results \cite{YXDuan2019}.

Constant photon energy $k$$_x$-$k$$_y$ FS mappings were taken to clarify FSs topological structures. Figs.\ 1(c) and 1(d) display the FS intensity maps measured at 18 K with 85 and 92 eV photon energies, respectively. FSs calculated at corresponding $k$$_z$ were overlaid. Most FSs agreed well with DFT calculations. This may possibly be due to the fact that FSs are contributed by In and Pt characters. Ce 4$f$ electrons may also participate in FS formation, but that contribution is relatively weak. Also, it can be found that Fermi sheets shape around $\Gamma$ points vary greatly with photon energy, while Fermi sheets shape at the BZ corner $M$ point changes little. This is consistent with the coexistence of 2D and 3D characters suggested by theoretical calculations \cite{YXDuan2019, BShen2017, Klimczuk2014}.

The 4$d$ $\rightarrow$ 4$f$ on-resonant ARPES measurements were performed with 123 eV photons to strengthen the Ce 4$f$ electron photoconduction matrix element. Figs.\ 2(a) and 2(b) compare off-resonance with 114 eV photons and on-resonance with 123 eV photons photoemission spectra at 20 K, respectively, by using $s$-polarized light. The off-resonance spectra is dominated by Pt 5$d$ and In 5$p$ states derived dispersive bands. They show a density of states close to the Fermi energy ($E$$_F$) of non-$f$ orbital character. The Ce 4$f$ state is enhanced in on-resonance data. This is also seen from the integral spectra in Fig.\ 2(d). Two nearly flat bands originate from the spin-orbit splitting of $f$$^1$ final state can be observed in the on-resonance data. This was seen in prior experiments \cite{YXDuan2019}. The $\sim$ 300 meV is assigned to 4$f$$^1_{7/2}$. The other, near $E$$_F$, is attributed to 4$f$$^1_{5/2}$ \cite{YXDuan2019}. Similar structures have been observed for other members of the Ce$_m$$M$$_n$In$_{3m+2n}$ family \cite{QYChen2017, QYChen2018A, QYChen2018B, QYao2019, Fujimori2003, Fujimori2006, HJLiu2019} and also in CeCu$_2$Si$_2$ \cite{SPatil2016}.

Fig.\,2(c) shows the on-resonance spectra measured with $p$-polarized light. Compared with $s$-polarization, the 4$f$$^1_{5/2}$ quasiparticle peak measured with $p$-polarization is sharper. The 4$f$$^1_{7/2}$ quasiparticle peak is less sensitive to polarization. Compared to $s$-polarization, the spectrum's overall intensity under $p$-polarization is reduced. Fig.\ 2(e) is a comparison of the calculated density of state (DOS) of the isostructural compounds {\CPI} and LaPt$_2$In$_7$. Only the 4$f$ occupation is varied. Compared to the La 4$f$ electrons in LaPt$_2$In$_7$ (blue line) which are far away from $E$$_F$, the Ce 4$f$ electrons (red line) are also above $E$$_F$, but some fragments leak below $E$$_F$. This explains why Ce 4$f$ state enhancement is observed in the on-resonance spectrum. It is not as significant as the increase measured previously in U 5$f$ electrons \cite{JQM_PRL2013}.

\begin{figure}[tb!]
\vspace*{-0.2cm}
\begin{center}
\includegraphics[width=0.95\columnwidth,angle=0]{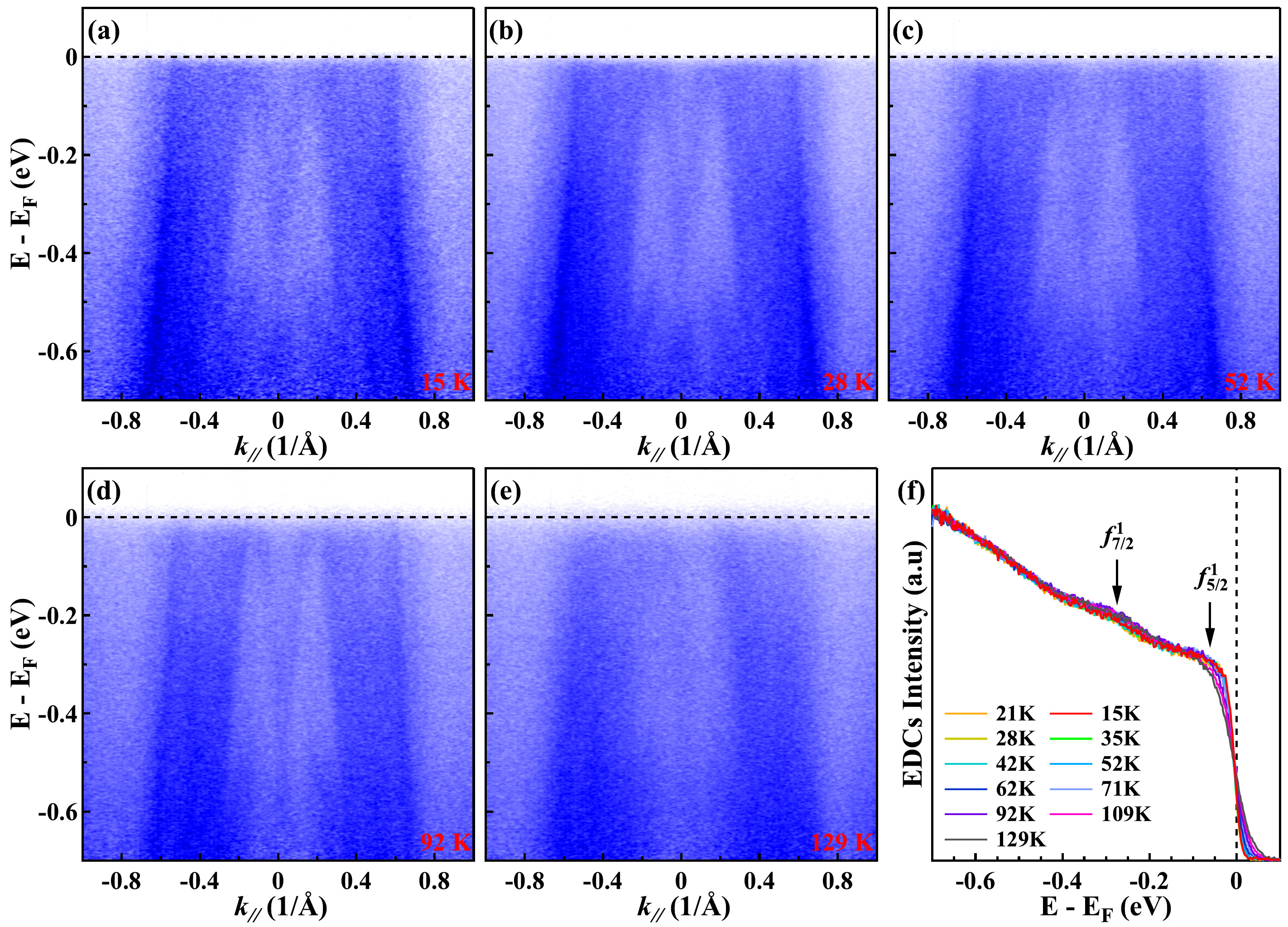}
\end{center}
\vspace*{-0.7cm}
\caption{(color online) The band structure along the A-Z-A direction of {\CPI}, measured with 85-eV photons at various temperatures. ({\bf {a}}) - ({\bf {e}}) Photoemission intensity plots at different experiment temperatures. ({\bf {f}}) Temperature dependence EDCs, which were integrated over the momentum range show in (a) - (e). Momentum cuts taken with 85 eV photons cross (0, 0, 16.84$\frac{2\pi}{c}$), close to $Z$, and thus labeled $A$-$Z$-$A$ for simplicity.}
\end{figure}

The $f$ electron properties of HF compounds, localized or itinerant, and how they transform as a function of temperature, have been critical issues in understanding HF underlying physics. Above a characteristic temperature, $T$*, $f$ electrons are completely localized. But well below $T$*, $f$ electron are described by the itinerant model. Below $T$*, $f$ electrons exhibit a dual nature. They are partially localized and partially itinerant \cite{YFYang2008, YFYang2016, Nakatsuji2004}. Temperature-dependent ARPES measurements on {\CPI} were conducted to better understand how Ce 4$f$ electron localization and itinerancy evolves with temperature.

Two photon energies were selected. One at 85 eV. The other is on-resonance 121 eV. 85 eV photons are not as good at detecting Ce 4$f$ electrons as on-resonance 121 eV, some information about Ce 4$f$ electrons can still be obtained. Figs.\ 3(a)-(e) display the conduction bands temperature evolution measured along $A$$^{\prime}$-$Z$$^{\prime}$-$A$$^{\prime}$ direction. The conduction bands show strongly dispersive features throughout the measured temperature range. Looking closely, we can see as temperature increases, conduction bands blur, but remain, like the V-shape band at the $Z$$^{\prime}$ point. Fig.\ 3(f) shows energy distribution curves (EDCs). They are integrated over the measured momentum range shown. As indicated by arrows, two expected weak structures originating from 4$f$$^1_{7/2}$ and 4$f$$^1_{5/2}$ states, respectively, have been observed. It can be observed that the 4$f$$^1_{7/2}$ state always exists, all the way up to 129 K. Further quantitative analysis of the 4$f$$^1_{5/2}$ electron will follow.

\begin{figure*}[tb]
\centering
\includegraphics[width=2\columnwidth,angle=0]{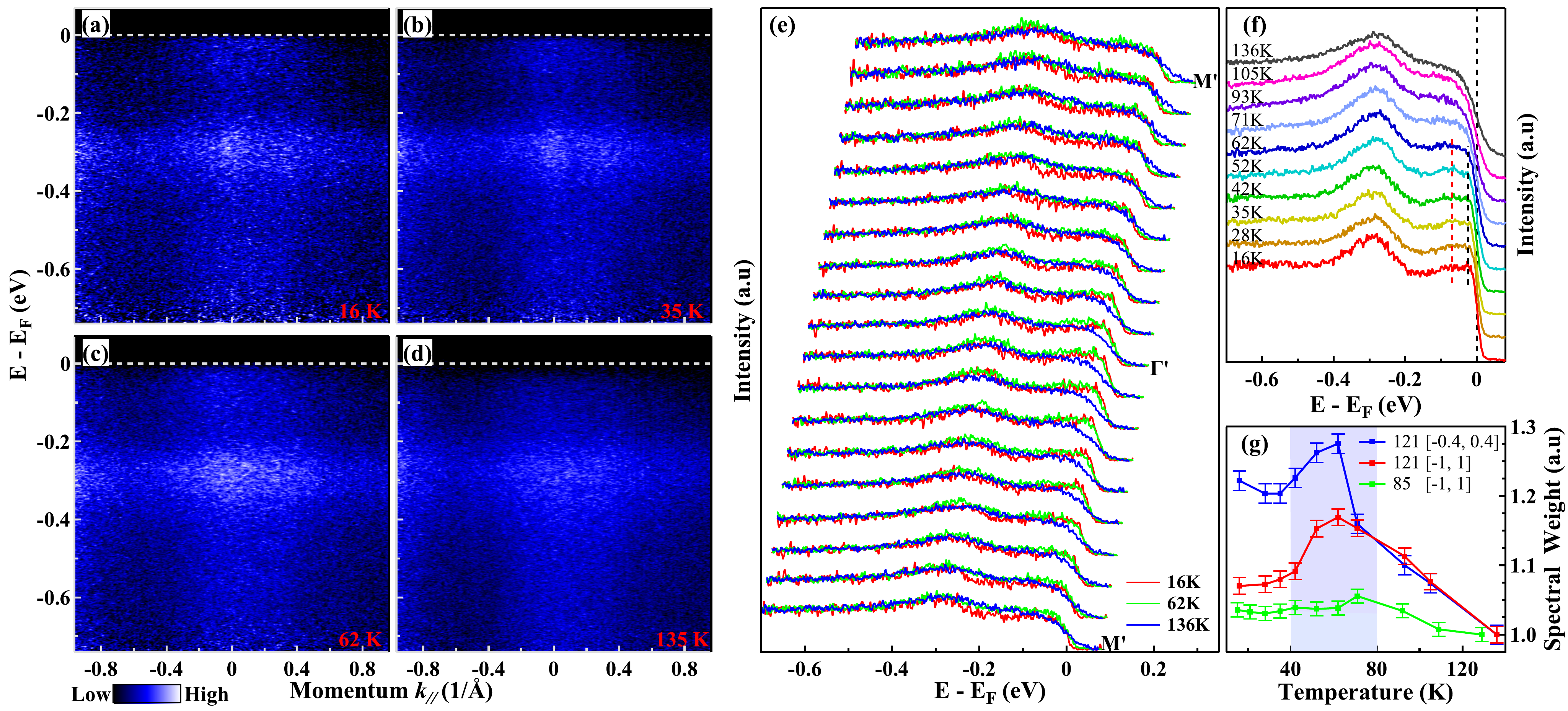}
\caption{(color online) Temperature evolution of the heavy quasiparticle band. ({\bf {a}}) - ({\bf {d}}) {\CPI} band structure measured along $M$$^{\prime}$-$\Gamma$$^{\prime}$-$M$$^{\prime}$ direction with 121 eV photons at different temperatures. ({\bf {e}}) Detailed ARPES spectra of {\CPI} measured at 16, 62, and 136 K. ({\bf {f}}) Angle-integrated photoemission spectroscopy at various temperatures. The integrated momentum range is [-0.4, 0.4]. ({\bf {g}}) $T$ dependence of the quasiparticle spectral weight near $E$$_F$, integrated over [$E$$_F$ - 100 meV, $E$$_F$ + 20 meV]. The momentum integral ranges for blue, red, and green lines are [-.04, 0.4], [-1, 1] and [-1, 1], respectively. The unit of momentum used here is ${\AA}$$^{-1}$. }
\end{figure*}

Figs.\ 4(a)-(d) display the temperature-dependent Ce 4$d \rightarrow 4f$ on-resonant ARPES measurements along the $M$$^{\prime}$-$\Gamma$$^{\prime}$-$M$$^{\prime}$ direction with 121 eV photons. Two flat bands with the Ce 4$f$ electron character exist throughout the measured temperature range. They are weakened at high temperatures. The intensity of the two flat bands is strongly momentum dependent. Fig.\ 4(e) shows a comparison of the EDCs measured at 16, 62, and 136 K. The main structure is similar. The structure near $E$$_F$ varies significantly with temperature. Quasiparticle peaks located near $E$$_F$ show a strong momentum and temperature dependence. Within a very small energy range very close to $E$$_F$, about 40 meV, EDCs intensities increase as temperature decreases. Over a slightly larger energy range, such as 100 meV, EDCs intensity evolution is no longer monotonous. Fig.\ 4(f) shows EDCs temperature evolution integrated between the momentum range [-1 {\AA}$^{-1}$, 1 {\AA}$^{-1}$]. Heavy quasiparticle band formation begins at a temperature $T$ much higher than its collective hybridization temperature $T$$^*$, which, for {\CPI} is 40 K \cite{Warren2011} or 20 K \cite{HSakai2012}. This is consistent with previous observations by other research groups made for other members of the family Ce$M$In$_5$ \cite{QYChen2017, QYChen2018A, QYChen2018B} and Ce$_2$PdIn$_8$ \cite{QYao2019}.

Fig.\ 4(g) quantitatively shows the $f$ electron spectral weight evolution with temperature. Spectral weight has been normalized to the highest temperature data. EDC intensity with different momentum integral widths exhibits similar temperature evolution behavior. These results show that $f$ electrons spectral weight defined in this energy range has an unexpected temperature evolution behavior. As cooling begins, the $f$ spectral weight increases as temperature decreases. Then, $f$ spectral weight decreases with further decrease of temperature. Similar behavior was found in the 85 eV data set (green line). $f$ spectral weight temperature evolution behavior is inconsistent with those of other research groups, although we believe that heavy quasiparticle bands are being formed already at elevated temperatures \cite{QYChen2017, QYChen2018A, QYChen2018B, QYao2019}. These results reveal that as the temperature falls, $f$ electrons transit from localized to more itinerant and back to more localized, as discussed further below. Relocalization has been reported in {\CPI} by Knight shift measurement and interpreted as a significant precursor to ordering antiferromagnetically \cite{Warren2011}. The relocalization temperature, 14 K, determined by Knight shift is much lower than the temperature observed here, $\sim$ 60 K.

Previous ultrafast experiments have reported that photoinduced transient reflectivity varies greatly around 60 K \cite{RYChen2016}. In-plane resistivity shows a bulge around 40 - 80 K \cite{VASidorov2013}. Looking closely at the EDCs in Fig.\ 4(f), as indicated by red and black dashed lines, discloses that CEF splitting begins in the 4$f$$^1_{5/2}$ band around 60 K. The effect becomes more obvious as temperature decreases. It appears that the CEF splitting of $f$ electrons near the $E$$_F$ is responsible for the relocalization. At low temperatures, a typical energy scale in a HF system is less than 10 meV. As energy resolution is not good enough and/or the peak of the low-lying heavy quasiparticle band is not sharp enough, no lowest-lying CEF excitation, less than 10 meV, near $E$$_F$ were observed. Some ARPES experiments which focused on the low-lying flat band near $E$$_F$, with a binding energy of about 7 meV (Ce$_2$IrIn$_8$) \cite{HJLiu2019} or 4 meV (YbRh$_2$Si$_2$) \cite{SKMo2012}, suggest that a coherent state developed below the characteristic temperature, $T$$^*$. Inelastic neutron scattering measurements suggest that the 4$f$ conduction electron interaction energy scale is comparable to that of low-energy CEF splitting \cite{TWillers2010, Christianson2002, Christianson2004}. Comparing it with the $f$ spectral weight defined by large-scale energy, the low-lying CEF excitation may better reflect the characteristic temperature $T$* of the HF compounds.

To summarize, the 3D electronic structure and 4$f$ electron behaviors of the HF superconductor {\CPI} were studied by ARPES over a wide temperature range. The results provide clear evidence that (i) FSs have coexistence of quasi-2D and 3D topology, which is consistent with DFT calculations; (ii) $f$ electrons begin to evolve into the formation of HF states at a temperature much higher than the characteristic temperature $T$*; and (iii) the CEF splitting, i.e., the low-lying heavy quasiparticle band caused by CEF, indicates the characteristic temperature $T$*.

This work was supported by the National Natural Science Foundation of China (Grant No.\ 11574402, No.11874264), ZDXKFZ (Grant No.\ XKFZ201703), the Innovation-driven Plan in Central South University (Grant No.\ 2016CXS032), the Natural Science Foundation of Shanghai (No.\ 17ZR1443300). This work was also supported through the Swedish Research Council (VR) and the Swedish National Infrastructure for Computing (SNIC), for computing time on computer cluster Triolith at the NSC center (Link{\"o}ping). Some preliminary data were taken at beamline 13U of the National Synchrotron Radiation Laboratory (NSRL).

{}
\end{document}